\documentclass[11pt,a4paper]{article}
\usepackage{epsf,feynmf,amsmath}
\usepackage{graphicx,color,mathrsfs,amsmath,amssymb}
\newcommand{\reels}{\mathbb{R}}
\def\RR{\hbox{\it I\hskip -2.pt R }}
\newcommand{\sla}{\not\!}
\title{ \bf The fate of the axial anomaly \\ in a finite field theory}
\author{Pierre Grang\'e $^a$, Jean-Fran\c cois Mathiot $^b$\thanks{e-mail: jean-francois.mathiot@clermont.in2p3.fr}\ \  and Ernst Werner $^c$ \vspace{0.3cm}\\
{\small \em  $^a$  Laboratoire Univers et Particules de Montpellier,CNRS/IN2P3, 
} \\ {\small \em Universit\'e Montpellier II, Place E. Bataillon F-34095 Montpellier Cedex 05, France} \vspace{0.3cm}\\
{\small \em  $^b$ Universit\'e Clermont Auvergne, Laboratoire de Physique de Clermont,
} \\ {\small \em CNRS/IN2P3, BP10448, F-63000 Clermont-Ferrand, France} \vspace{0.3cm}\\
{\small \em  $^c$  Institut f$\ddot u$r Theoretische Physik, Universit$\ddot a$t Regensburg,
} \\ {\small \em Universit$\ddot a$tstrasse 31, \ D-93053 Regensburg,  Germany}  }
\date{}
\begin{document}
\maketitle
\bibliographystyle{unsrt}
\abstract{The conservation of the vector current and the axial anomaly responsible for the $\pi_0 \to \gamma \gamma$ decay amplitude  are obtained in leading order within the Taylor-Lagrange formulation of fields considered as operator-valued distributions. As for gauge theories, where this formulation eliminates all divergences and preserves gauge-symmetry, it is shown that the different contributions can be evaluated directly in $4$-dimensional space-time,
with no restrictions whatsoever on the four-momentum of the internal loop, and without the need to introduce any additional non physical degrees of freedom like Pauli-Villars fields. We  comment on the similar contributions responsible for the decay of the Higgs boson into two photons.}
%
\section{Introduction} \label{intro}
The achievements of the Standard Model  of particle physics  in the {\it ab-initio} calculation of physical observables in the energy range accessible to present accelerators may be, at first sight, attributed to its formulation in terms of quantum field theory (QFT). It relies on the choice of a local Lagrangian constructed from the products of fields or their  derivatives at the same point, with well known properties like gauge  and Poincar\'e invariances.  However beyond the tree level approximation,  where additionnal divergent contributions show up, a more fundamental understanding is necessary. In the past, two attitudes appeared, different in their rationals. The very-first one adopts a pragmatic regularization,{\it ``\`a posteriori''}, of the loop contributions,  via for instance a na\"ive cut-off in momentum space. The second ones are more fundamental in the sense that they aim at working, from the very beginning, with a (mathematically) well-defined local Lagrangian, and lead to {\it ``\`a priori''} regularization procedures. Among the latters, the well known dimensional regularization ($DR$) operates through the definiton of the Lagrangian density in a space-time of dimension $D=4-\epsilon$.

 From a theoretical point of view, the regularization procedures should meet the requirements of gauge invariance and, in the first place, account  for elementary observed processes such as  the axial anomaly responsible for the $\pi_0 \to \gamma \gamma$ decay amplitude (the so-called triangular diagrams). These two constraints are not at all trivial to verify, depending on the choice of the regularization procedure. The use of a na\" ive cut-off, for instance, is known to violate gauge invariance.  Appropriate contributions to correct for this violation are then necessary. The  origin of this violation is well-known. It comes from the ill-defined behavior of elementary amplitudes corresponding to the triangular diagrams, prohibiting  any na\" ive change of variable in the relevant diverging integrals \cite{ryder}. This clear drawback of the use of a na\" ive cut-off can be removed by considering the Pauli-Villars regularization procedure. This one amounts to correct elementary propagators by enough subtractions with auxiliary unphysical masses $M_i$,  thereby removing the divergences of the elementary amplitudes. While this procedure does preserve gauge invariance and is able to reproduce the right axial anomaly \cite{pauli}, it relies on the use of auxiliary unphysical fields with ultimately infinite masses. This breaks, from the very beginning, the conservation of the axial current at tree level, for massless physical fields. 

At variance, the use of $DR$ respects gauge invariance. It is so  by construction of the  initial local Lagrangian  at $D=4-\epsilon$ dimensions ({\it ``\`a priori''} regularization procedure). The calculation of any axial transition in $DR$ necessitates nonetheless to choose, by hand, an adequate definition of $\gamma_5$ for $\epsilon \neq 0$, together with its anti-commutation relations with the other $\gamma$ matrices and the construction of the Levi-Civita tensor \cite{collins}. While such solution does indeed exist, it is  not unique and does not follow from any physical nor mathematical first principles. Its legitimity relies only on its ability to reproduce the physical observables under consideration. 

From the late nineties until recently,  a number of important studies have dealt with these problematics occuring throughout the very foundations of QFT. Some of the most recent works we shall refer to in the sequel are  concerned with regularization and renormalization issues and their aim to construct finite field theories \cite{EG,Scharf,prop1,prop2,GW_mass,GW}. The (mathematical) need to start from the outset with well defined field operators and Lagrangians open completely new perspectives for quantum field theories.  Common to these first principle attempts is the emphasis on new rigorous mathematical perequisites, the need to stick to the physical world, and an operational strategy in order to perform practical calculations of physical observables.

A way to reach these goals is the recently proposed Taylor-Lagrange regularization scheme ($TLRS$) by two of the authors of the present study \cite{GW_mass,GW}. It is genuine in that it takes into account the nature of quantum fields as operator valued distributions ($OPVD$) and operates directly in four dimensions.  In this scheme, the physical quantum fields are  defined as functionals of test functions with adequate boundary conditions. It is our purpose in this study to show how the early pathological behaviors mentioned above can be cured quite naturally in $TLRS$ through the fundamental definition of quantum fields as $OPVD$. Interestingly enough, we shall recall in this study how Pauli-Villars type subtractions do indeed occur within $TLRS$ without the need to consider extra non-physical degrees of freedom.

 The plan of our study is as follows. We recall in Sec.~\ref{TLRS} the main properties of $TLRS$. We calculate in Sec.~\ref{triang} the triangular diagrams and check the conservation of the vector current and calculate the axial anomaly. Final remarks are drawn in Sec.~\ref{conc}. We recall in  \ref{taylor} the  calculation of elementary amplitudes within $TLRS$ in the $UV$ as well as $IR$ domains. The explicit calculation of the relevant non-trivial integrals  is done in  \ref{RI}.

\section{The Taylor-Lagrange regularization scheme} \label{TLRS}
As emphasized long ago (Refs.~\cite{bogo}-\cite{haag}, see also Refs.~\cite{schw,tkac}), any field $\phi(x)$ - taken here as a scalar field for simplicity - should be considered as an $OPVD$. The first derivation of the renormalization group equations refers indeed to this unique property \cite{stue}. It is a mathematical necessity for a sound  extension, in the sense of distributions, of the otherwise ill-defined product of two fields at the same space-time point \cite{HORM1, HORM2}. However, it was recognized only much later how to deal in practice with these requirements \cite{EG,Scharf, prop1,prop2}. Here, for the purpose of a self-contained presentation, we  recall the proposed procedure called $TLRS$. This mathematical approach to the distributional extension of singular distributions is detailed in Ref.~\cite{GW}. First applications for calculations in light-front dynamics are given in Refs.~\cite{LFD,LFD_PV}, for gauge theories in Ref.~\cite{GW_gauge} and for the calculation of the radiative corrections to the Higgs mass, within the Standard Model, in Ref.~\cite{higgs}.
 
According to the theory of distributions \cite{schwartz} (previously known also as generalized functions \cite{gene}), the distribution $\phi$ defines a functional $\Phi$ with respect to a test function $\rho$. This functional can be written as 
\begin{equation}
\Phi(\rho) \equiv \int d^4y \phi(y) \rho(y).
\end{equation}
Due care is needed in the interpretation of this functional, for the  distribution $\phi(y)$ is not to be looked at its point values but at its action on the set of test functions $\rho$ belonging to the space of rapidly decreasing functions, the so-called Schwartz's space $\mathscr{S}$ (see \ref{taylor}).
This functional is a linear continuous form on $\mathscr{S}$, hence an element of its topological dual $\mathscr{S}^{\prime}$, that is a {\it density  measure} $d\mu_{\rho}(y)=d^4y\rho(y)$ in the sense of Measure Theory \cite{schwartz,Integ}. 
We recall here that if $P_i$ is the paved-support of $\rho(y_i)$ then $\Phi(\rho)$ may be  defined as 
\begin{equation}
\Phi(\rho) \equiv \mu(\rho) \equiv \int_{\RR^{\!\!^4}}\phi(y)d\mu_{\rho}(y)\doteq{\displaystyle\sum_{P_i}} \phi(y_i)\mu_{\rho}(P_i),\nonumber
\end{equation}
for any $y_i$ belonging to $P_i$. 
The physical field we are interested in, denoted by $\varphi(x)$, is then constructed from the translation of $\Phi(\rho)$ defined  by 
\begin{equation} \label{conv}
\varphi(x)  \equiv \int d^4y \phi(y) \rho(\vert x-y \vert),
\end{equation}
where the reflection symmetric test function $\rho$ belongs to  $\mathscr{S}$. This convoluted definition insures the existence of both $\varphi$ and its Fourier transform as continuous functions - as well as all their derivatives - over their respective spaces. It is easy to check that  $\varphi$ is also solution of the initial equation of motion, here the Klein-Gordon equation. Hence, in any case, $\varphi$ is a {\it bona fide}  physical field.   

The physical interest to use the test function $\rho$ is to smear out the original distribution in a  space-time domain of typical extension $a$, the size of the paved-sector $P_i$ defined above. The test function can thus be characterized by $\rho_a(x)$ and the physical field by $\varphi_a(x)$. Requiring locality for the initial Lagrangian in terms of $\varphi$ implies considering the subsequent  limit $a \to 0$,\,herafter dubbed {\it the continuum limit.} In this process, the scaling properties ought to be preserved 
\begin{equation} \label{scale}
\rho_a(x) \to \rho_\eta(x) \ ; \  \varphi_a(x) \to\varphi_\eta(x),
\end{equation}
where $\eta$ is an arbitrary, dimensionless, scaling variable (called also scale in the following) since in the limit $a \to 0$, we also have $a/\eta \to 0$, for any $\eta >1$. This arbitrary scale just governs the way the continuum limit is reached. Any physical observable should of course be independent of this dimensionless regularization scale, as emphasized in Ref.~\cite{jfm}.

For practical calculations, it is convenient to construct the physical fields in momentum space. If we denote by $f_\eta$ the Fourier transform of the test function $\rho_\eta$, we can write ${\varphi_\eta}$ in terms of creation and annihilation operators, leading to \cite{GW}
\begin{equation} \label{fk}
\varphi_\eta (x)\!=\!\!\int\!\frac{d^3{\bf p}}{(2\pi)^3}\frac{f_\eta(\varepsilon_p^2,{\bf p}^2)}{2\varepsilon_p}
\left[a^\dagger_{\bf p} e^{i{p.x}}+a_{\bf p}e^{-i{p.x}}\right],
\end{equation}
with  $\varepsilon^2_p = {\bf p}^2+m^2$.  
Each propagator being the contraction of two fields is proportional to $f_\eta^2$. 

The test function $f_\eta$ in momentum space  is a dimensionless quantity. It should therefore be expressed in terms of dimensionless arguments. To do that, we shall introduce a fixed scale $M_0$ to "measure" all momenta. The choice of $M_0$ is arbitrary. In practical calculations, $M_0$ can be any of the non-zero physical mass of the theory under consideration. Physical observables should of course be independent of this scale. As we shall see below, a change of $M_0$ is just equivalent to a redefinition of $\eta$. This scale is reminiscent of the definition of the unit of mass  $\mu$ in $DR$, with the identification of the dimensionless regularization scale $\eta$  by $\mu=\eta M_0$, as argued in Ref.~\cite{jfm} (see also Refs.~\cite{GW_mass,GW}).
Note that the condition $a \sim 0$ implies, in momentum space, that $f_\eta$  is constant almost everywhere. It is easy to see that  such constant can be chosen equal to $1$ in order to conserve the normalization of the fields, without any loss of generality. 

The function $f_\eta$ belongs also to the space $\mathscr{S}$, with infinite support. To construct it from a practical point of view, we shall start from a sequence of functions, denoted by $f_\alpha$, with compact support, and build up as partition of unity \cite{GW}. This construction is universal, and refers to the locally finite open covering - with their subordinate partition of unity \cite{PU1,PU2,PU3} - of the Minkowskian and Euclidian manifolds as paracompact entities. This function is equal to $1$ almost everywhere and is $0$ outside a finite domain of $\reels^{+ 4}$, along with all its derivatives (super-regular function) \footnote{Note that if $f$ is a partition of unity, $f^2$ is also a partition of unity and we have $f^2 \sim f \sim1$ in the continuum limit.}. 
The parameter $\alpha$, chosen for convenience between 0 and 1, controls the lower and upper limits of the support of $f_\alpha$.  The limit $\alpha \to 1^-$ defines the continuum limit for which $f \sim 1$. A particular example of the construction of $f_\alpha$ can be found in Ref.~\cite{LFD}.

We can thus write schematically any (one-loop) amplitude associated to a singular distribution $T(X)$ as
\begin{equation} \label{cala}
{\cal A}_\alpha = \int_0^\infty dX \ T(X) \ f_{\alpha}(X),
\end{equation}
for a dimensionless variable $X$. We shall consider here for the purpose of this study a one-dimensional variable only. The general case is discussed in Ref.~\cite{GW}.  
We must now verify that taking the  continuum limit $\alpha \to 1^-$ for the test function $f_\alpha$ does lead to a well defined, finite, amplitude. 
For that, we shall verify that the amplitude 
\begin{equation} \label{Aeta}
A_\eta \equiv \lim_{\alpha \to 1^-}A_\alpha
\end{equation}
 is independent of the upper and lower boundaries of the support of the test function $f_\alpha$, denoted by  $X_{min}$ and $ X_{max}$, respectively. It is easy to see that with a na\" ive construction of $f_\alpha$, using a sharp cut-off at $X_{max}$ for instance, this condition is not verified.
In order to preserve the scaling properties (\ref{scale}), we  shall consider a running boundary $H_\alpha(X)$ defined by
\begin{equation} \label{running}
f_\alpha(X \ge H_\alpha(X)) = 0,
\end{equation}
with
\begin{equation} \label{scaling}
H_\alpha(X)\equiv \eta^2 X g_\alpha(X) + (\alpha - 1),
\end{equation}
where $\eta$ is an arbitrary dimensionless scale\,\footnote{The square of $\eta$ in (\ref{scaling}) is just for convenience since $X$ is in general constructed from the square of a momentum.} which should only be larger than $1$.  Remarkably enough, this condition defines at the same time the $UV$ and $IR$ boundaries once $f$ is constructed from a partition of unity. The constant term in the {\it r.h.s.} of Eq.~(\ref{scaling}) is only relevant in the $IR$ limit.
The function $g_\alpha(X)$ is chosen so that when $\alpha \to 1^-$, $X_{max}$ defined by 
\begin{equation}
X_{max}= H_\alpha(X_{max}),
\end{equation}
tends to infinity. A typical (but not unique) simple form for $g_\alpha(X)$ is given by 
\begin{equation} \label{gal}
g_\alpha(X)=X^{\alpha-1},
\end{equation}
with $0 < \alpha < 1$.
In the limit $ \alpha \to 1^-$, we have $X_{max} \to \infty$ with $g_\alpha (X) \to 1^-$ except in the asymptotic region  $X \sim X_{max}$. In the $IR$ domain, we have similarly $X_{min} \to 0$ in this limit. The explicit form for $X_{min}$ and $X_{max}$ can be found in Ref.~\cite{LFD}.
The conditions (\ref{running},\ref{scaling},\ref{gal}) amount to an ultra-soft long-range behavior, {\it i.e.} an infinitesimal drop-off of the test function in the asymptotic region, with a drop-off rate governed by the arbitrary scale $\eta$.

One may argue that the use of test functions is, by construction, equivalent to a na\" ive regularization method. As mentioned above, using a  step function for the test function, one recovers from (\ref{cala}) a usual {\it brute force} regularization with a cut-off in momentum space. In this case, however, the continuum limit cannot be reached at the level of each individual amplitude, as required by (\ref{Aeta}). The regularization induced by the test function in $TLRS$, with its boundary condition (\ref{running}) and the scaling property (\ref{scaling}), is of a different nature. We shall call it {\it intrinsic} as opposed to an {\it extrinsic} regularization in the case of a cut-off (or using Pauli-Villars regularization or $DR$ as well). For  an {\it extrinsic} regularization scheme, infinities are recovered for any individual amplitude in physical conditions, {\it i.e.} for an infinite cut-off or in 4-dimensional space. This is not so for an {\it intrinsic} regularization procedure, like $TLRS$, where all individual amplitudes are finite in 4-dimensional space-time and with no restrictions whatsoever on available momenta, hence the denomination of finite field theory. 

From the construction of $TLRS$ based on the $OPVD$ nature of quantum fields, together with the scaling properties of the test function in the $UV$ domain (continuum limit), any elementary amplitude is thus finite in the $UV$ as well as $IR$ domains. This is a unique property based on the construction of the test function as a partition of unity. {\it The extension of any singular operator (see \ref{taylor}) is thus well defined, but depends on an arbitrary dimensionless scaling variable $\eta$}. As a consequence, all bare parameters are finite by construction - in physical $D=4$ conditions with no {\it ad-hoc}  subtractions in the $UV$ as well as $IR$ domains. In this sense, $TLRS$ is at the same time a regularization procedure as well as a renormalization scheme, with the same acronym $TLRS$. Physical observables are thus made independent of the scaling variable $\eta$ by a finite renormalization, as it is the case in any interacting many-body theory.

The explicit calculation of the elementary amplitudes $A_\eta$ in the $UV$ as well as $IR$ domains is recalled in \ref{taylor}. From a purely practical point of view, the use of $TLRS$ in the $UV$ domain is expected to give similar results as $DR$ with the minimal subtraction ($MS$) renormalization scheme. Indeed, while poles in $1/\epsilon$ are removed by hand in $DR+MS$, they are absent by construction in $TLRS$. Both schemes lead also to mass-independent coefficients of the renormalization group equation since they refer to the scaling behavior of the amplitudes in the $UV$ domain, with no other explicit arbitrarily large mass-scale. Both are also {\it ``a-priori''} regularization procedures. However, the natural extension of singular operators in the $IR$ domain, in terms of Pseudo-Functions as recalled in  \ref{IR}, is a unique property of $TLRS$. It enables in particular to tackle from a well defined mathematical framework the behavior of any amplitude in the massless limit. The calculation  of the  two-point function of the  scalar field \cite{GW_mass} in a perturbative expansion in terms of the mass-parameter, or the taming of the $IR$ singularities of the gauge field propagator in the light-cone gauge \cite{GW_gauge} are particularly suggestive examples of how $TLRS$ should be used in presence of such singularities.

\section{Calculation of the triangular diagrams} \label{triang}
\begin{figure}[b]
\centerline{\includegraphics[width=25pc]{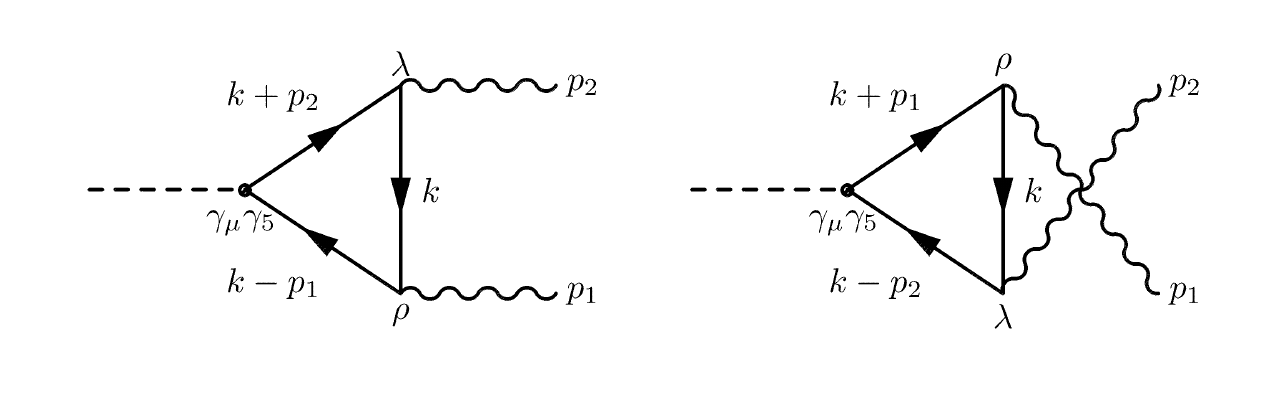}}
\begin{center}
\caption{Triangular diagrams.  \label{fig1}}
\end{center}
\end{figure}

The amplitude corresponding to the transition of an axial current into two photons is indicated, in leading order, in Fig.~\ref{fig1}. The first diagram is given by
\begin{equation}
T_{\lambda\rho\mu}=-\lim_{\alpha \to 1^-}\int \frac{d^4k}{(2\pi)^4}\frac{I_{\lambda\rho\mu}}{(k^2-m^2)\left[(k-p_1)^2-m^2\right]\left[(k+p_2)^2-m^2\right]}F_\alpha(k,p_1,p_2),
\end{equation}
with
\begin{equation}
I_{\lambda\rho\mu}=Tr\left[\gamma_\rho(\sla k+m)\gamma_\lambda (\sla k + \sla p_2 +m)
\gamma_\mu \gamma_5 (\sla k - \sla p_1 +m)\right].
\end{equation}
The fermion mass is denoted by $m$. The second diagram is derived from the first one by the replacement $p_1 \to p_2, \rho \to \lambda$. The factor  $F_\alpha$ accounts for the presence of the test functions attached to each individual fermion propagator (with the replacement $f^2_\alpha \to f_\alpha$). It is given by
\begin{equation}
F_\alpha=f_\alpha \left[\frac{k^2}{M_0^2}\right] f_\alpha \left[\frac{(k-p_1)^2}{M_0^2}\right] f_\alpha \left[\frac{(k+p_2)^2}{M_0^2}\right].
\end{equation}
The calculation is done, in $TLRS$, at $D=4$ so that there is no ambiguity whatsoever on the definition of $\gamma_5$. We  get immediately 
\begin{equation}
T_{\lambda\rho\mu}=-2\ \lim_{\alpha \to 1^-}\int_0^1 dx \int_0^{1-x} dy \int \frac{d^4K}{(2\pi)^4}
\frac{I_{\lambda\rho\mu}}{(K^2-\Delta^2)^3}F_\alpha(K-Q,p_1,p_2),
\end{equation}
with $K=k+Q$ and 
\begin{equation}
Q=y p_2-x p_1\ ,\ \Delta^2=m^2-x y q^2\ ,\ q=p_1+p_2.
\end{equation}
Note that any change of variable in the calculation of the integrals is legitimate in $TLRS$ since any elementary amplitude is finite thanks to the presence of the test functions (and hence the argument $K-Q$ in $F_\alpha$). The calculation of $T_{\lambda\rho\mu}$ proceeds further through the evaluation of the following four different integrals given, in Euclidian space, by
\begin{subequations}\label{inte}
\begin{eqnarray}
J_0&=& \lim_{\alpha \to 1^-}\int \frac{d^4{\bf K}}{(2\pi)^4}\frac{1}{({\bf K}^2+\Delta^2)^3}F_\alpha({\bf K-Q},{\bf p_1},{\bf p_2}),
\label{intea}\\
J_2&=& \lim_{\alpha \to 1^-}\int \frac{d^4{\bf K}}{(2\pi)^4}\frac{{\bf K}^2}{({\bf K}^2+\Delta^2)^3}F_\alpha({\bf K-Q},{\bf p_1},{\bf p_2}),
\label{inteb}\\
J_1^\mu &=& \lim_{\alpha \to 1^-}\int \frac{d^4{\bf K}}{(2\pi)^4}\frac{{\bf K}^\mu}{({\bf K}^2+\Delta^2)^3}F_\alpha({\bf K-Q},{\bf p_1},{\bf p_2}),
\label{intec}\\
J_2^{\mu\nu} &=&\lim_{\alpha \to 1^-}\int \frac{d^4{\bf K}}{(2\pi)^4}\frac{{\bf K}^{\mu}{\bf K}^{\nu}}{({\bf K}^2+\Delta^2)^3}F_\alpha({\bf K-Q},{\bf p_1},{\bf p_2}).
\label{inted}
\end{eqnarray}
\end{subequations}
The first integral is finite in absence of the test functions. The continuum limit $\alpha \to 1^-$ is thus straightforward, with $F_\alpha \to 1$. We thus get
\begin{equation}
J_0=\frac{1}{(4\pi)^2}\ \frac{1}{2\Delta^2}.
\end{equation}
The three other integrals should be calculated with  due account of the test functions. Their calculation is detailed in \ref{RI}. We get, in the continuum limit,
\begin{subequations}\label{intef}
\begin{eqnarray}
J_2&=& \frac{1}{(4\pi)^2}\ \mbox{Log}\left[\eta^2\frac{m^2}{\Delta^2}\right],\\
J_1^\mu &=& 0,\\
J_2^{\mu\nu} &=&\frac{1}{4}\delta^{\mu\nu} \left( \Delta^2 J_0+J_2\right) .
\end{eqnarray}
\end{subequations}
The integral $J_2$ is expressed in terms of the dimensionless regularization scale $\eta$ introduced in the previous section{\footnote{We have chosen here $M_0\equiv m$.}}. Note that we recover, in $TLRS$, the standard expression for $J_2^{\mu\nu}$ as calculated in $DR$, although our calculation is done at $D=4$. This result is entirely due to the presence of the test functions in (\ref{inte}), as shown in \ref{RI}. We recover also rotational invariance which results in $J_1^\mu=0$. Note that these results are only valid in the continuum limit. 

With the results (\ref{intef}), it is straightforward to check the  conservation of the vector current:
\begin{equation}
p_1^\rho\ T_{\lambda\rho\mu}=p_2^\lambda\ T_{\lambda\rho\mu}=0.
\end{equation}
For the axial current in the massless limit, we get
\begin{equation}
q^\mu\ T_{\lambda\rho\mu}=-\frac{1}{2\pi^2}\varepsilon_{\lambda\rho\alpha\beta}\  p_1^\alpha \ p_2^\beta.
\end{equation}
We recover here two known important features: gauge invariance is preserved through the conservation of the vector current, while the axial current is broken in the massless limit, as demanded by the axial anomaly, with the right contribution \cite{ryder}. It is remarkable that these two features do emerge naturally when using $TLRS$ in physical conditions, {\it i.e.} using four-dimensional space-time from the very beginning, with no additional non-physical degrees of freedom nor any cut-off in momentum space, nor any ad-hoc subtraction. 

\section{Final remarks} \label{conc}
The calculation of the triangular diagrams is a very powerful  testing ground in order to understand the main properties of any regularization procedure. It should of course provide a finite, regularization scale independent, contribution since this loop contribution does appear in leading order, {\it i.e.} with no tree level contribution. Moreover, it should  preserve gauge invariance, and provide at the same time the right anomalous contribution to the axial-vector-vector transition in order to account for the $\pi_0 \to \gamma \gamma$ decay amplitude.

While the calculation of these diagrams is by now a text-book exercise for many of the standard regularization procedures, we investigated in this study how to recover these properties within the recently proposed $TLRS$. This regularization procedure does provide a very clear and coherent picture. All the known properties, like gauge invariance and the right axial anomaly, do appear naturally in this scheme. This is a direct consequence of the fundamental properties of $TLRS$.  As an ``{\it a-priori}'' regularization procedure, it provides a well defined mathematical meaning to the local Lagrangian we start from in terms of products of $OPVD$ at the same point. It also yields a well defined unambiguous strategy for the calculation of elementary amplitudes which are all finite in strictly 4-dimensional space-time and with no new non-physical degrees of freedom nor any cut-off in momentum space  - as an``{\it intrinsic}'' regularization procedure. 

Note that the use of $TLRS$ shed also some light on similar calculations of one-loop contributions in leading order, like for instance those responsible for the $H \to \gamma \gamma$ transition, where $H$ is the so-called Higgs boson. With the calculation of the relevant integrals in Eqs.~(\ref{intef}), we recover immediately the known results obtained using $DR$ \cite{Hgg} or PV-substractions \cite{UlfM}. Although loop contributions are finite even in the absence of a regularization procedure, as noticed in Ref.~\cite{WWW}, we emphasize the mathematical necessity to start with a well defined Lagrangian, ${\it i.e.}$ to use anyhow an ``{\it a-priori}'' regularization procedure from the very beginning. This can be done using $DR$ or $TLRS$.

The benefits of well-defined Lagrangians go far beyond the present study in so far as the overall balance of $IR$- and $UV$-divergences conditions the reliability of a model. In this sense, the study of $TLRS$ as a unique regularization/renormalization scheme which starts from a well defined Lagrangian (``{\it a-priori''} regularization scheme) together with a well defined strategy to calculate any elementary amplitude which is finite by construction in physical, $D=4$, conditions (``{\it intrinsic}'' regularization scheme) is of particular interest. Contrarily of many, if not all, other regularization procedures, $TLRS$ does exhibit a fundamental {\it dimensionless} scaling variable, called $\eta$, without any other new mass scale. This should enable a completely new investigation of the properties of $QFT$ in the massless limit.

\appendix
\section*{Appendix}
\section{Extension of singular operators} \label{taylor}
\subsection{Extension in the ultra-violet domain} \label{ext}
The test function should belong to the Schwartz's space $\mathscr{S}$ of rapidly-decreasing function. From a general point of view, a function $\rho$ of the four-dimensional vector ${\bf x}$, with $r^2\!\!=\!\!{\displaystyle\sum_{i=1}^{4}}(x_i)^2$,  is of rapid-decrease if for any power $\alpha\!=\!\{\alpha_1...\alpha_4\}$ and for any order $\beta_i$ with 
$\mid\!\!\beta\!\!\mid\,=\!{\displaystyle\sum_{i=1}^{4}}\beta_i$, it verifies 
\begin{equation}
\displaystyle\lim_{r \rightarrow \infty}\!\mid\!\!x^\alpha\!\partial^\beta\!\rho(\bf{x})\!\!\mid\rightarrow\!0, \mbox{with}\ \ 
x^\alpha\!\!=\!\!{\displaystyle\prod_{i=1}^{4}}{x_i}^{\!\!\alpha_i}\ \mbox{and}\  \partial^\alpha\!\!=\!\!\frac{\partial^{\mid\alpha\mid}}{{\!}_{\partial{{{x_1}^{^{\!\!\!\!\!\!\alpha_1}}}}...\partial{{x_4}^{^{\!\!\!\!\!\!\alpha_4}}}}}.\nonumber
\end{equation}
The rapidly-decreasing test function is then equal to its Taylor remainder of any order $k$. We can thus apply Lagrange's formula to $f_\alpha$, after separating out an intrinsic scale $\lambda$ from the dynamical variable $X$, with the reduced singular distribution $\overline T(X)$ defined by
\begin{equation}
\overline T(X)=\lambda T(\lambda X).
\end{equation}
In order to keep track throughout our derivation of the scaling invariance embedded in the running condition (\ref{scaling}), Lagrange's formula should be applied to the test function, at fixed support \cite{LFD}. To do that, we shall write
\begin{equation}
f_\alpha(X)=f(X,H_\alpha(X))=f(X,H_\alpha(Y))\vert_{Y=X}
\equiv F_\alpha(X,Y)\vert_{Y=X}.
\end{equation}
We thus apply Lagrange's formula to the $X$-dependence of $F_\alpha$, and get:
\begin{equation} \label{faX}
F_\alpha(\lambda X,Y)=-\frac{X}{\lambda ^k k!}\int_\lambda ^\infty\frac{dt}{t} (\lambda -t)^k 
\partial_X^{k+1}\!\left[ X^k F_\alpha(Xt,Y)\right].
\end{equation}
The integral over $t$ is bounded from above by the support of the test function, and defined by $H(X)$ according to (\ref{running}) \cite{GW_mass,GW,LFD}. We thus have
\begin{equation}
Xt \leq H(X),
\end{equation}
so that with (\ref{scaling}) in the UV region
\begin{equation}
t \leq \eta^2 g_\alpha(X).
\end{equation}
In the continuum limit, with $g_\alpha \to 1$, we  get
\begin{equation}
t\leq \eta^2.
\end{equation}
Lagrange's formula (\ref{faX}) is valid for any order $k$, with $k\ge0$.
Starting from the general amplitude $A_\alpha$ written in (\ref{cala}),  and after integration by part, with the use of (\ref{faX}), we get
\begin{equation} \label{afin}
{\cal A}_\alpha = \int_0^\infty dX \ \widetilde T^>_\alpha (X) f_\alpha(X).
\end{equation}
In the continuum limit $f_\alpha \to 1$, {\it  i.e.} for $\alpha \to 1^-$, we have \cite{GW} 
\begin{equation}
\widetilde T^>_\alpha (X) \to \widetilde T^>_\eta (X)
\end{equation}
with
\begin{equation} \label{Tex}
\widetilde T_\eta^>(X)\equiv\frac{(-X)^{k}}{\lambda^k k!} \partial_X^{k+1} \left[ X \overline T(X)\right] \int_\lambda^{\eta^2} \frac{dt}{t} (\lambda-t)^k.
\end{equation}
This is the so-called extension of the singular distribution $T(X)$ in the $UV$ domain. The value of $k$ in (\ref{Tex}) corresponds to the order of singularity of the original distribution $T(X)$ \cite{GW}. In the continuum limit,  the integral over $t$ is independent of $X$ with the choice (\ref{running}) of the running boundary,  while the extension of $T(X)$ is  no longer singular due to the derivatives in (\ref{Tex}). We can therefore safely perform the limit $\alpha \to 1^-$ in (\ref{afin}), and get
\begin{equation}
{\cal A}_\eta = \int_0^\infty dX \ \widetilde T_\eta^>(X).
\end{equation}
In this equation, ${\cal A}_\eta $ is a well defined quantity which depends however on the arbitrary dimensionless scale $\eta$, as expected from general scaling arguments. This scale is the only remnant of the presence of the test function. 
Note that we do not need to know the explicit form of the test function in the derivation of the extended distribution $\widetilde T^>_\eta(X)$ in the continuum limit. 
We only rely on its mathematical properties. 

It is interesting to note that the physical amplitude (\ref{afin}) can be transformed alternatively in order to exhibit a $PV$-type subtraction \cite{LFD_PV}. With a typical  distribution $T(X)=\frac{1}{X+\lambda}$ with an intrinsic scale  $\lambda$, we can simply use Lagrange's  formula in the form
\begin{equation} \label{lat}
F_\alpha \left[\lambda X,Y \right]=-\int_\lambda^\infty dt \ \partial_t F_\alpha\left[X t,Y\right] \ ,
\end{equation}
we can rewrite the physical amplitude $\cal A_\alpha$ in the following way, after the change of variable $Z=Xt$ and  in the limit $\alpha \to 1^-$
\begin{equation}
{\cal A_\eta}=-\int_0^\infty dZ  \int_\lambda^{\eta^2} dt \  \partial_t \left[ \frac{1}{t}\ \overline T\left( \frac{Z}{t} \right) \right] \ .
\end{equation}
One thus gets immediately
\begin{equation}
\widetilde T^>(Z) = \frac{1}{Z+\lambda} - \frac{1}{Z+\eta^2}\ .
\end{equation}
We recover here a Pauli-Villars type subtraction, with an arbitrary dimensionless scaling variable $\eta^2>1$. Hence it should not be relentlessly large  contrary to Pauli-Villars masses which should tend to infinity in the final run. Moreover, the same calculation performed in Minkowskian space-time shows\,\cite{GW}\,that it is the cancelation of causality-violating time-like contributions coming from the two propagators which leaves a final and finite space-like contribution function of $\eta^2$. 

\subsection{Extension in the infra-red domain} \label{IR}
The extension of singular distributions in the $IR$ domain can be done similarly\,\cite{GW}. For an homogeneous distribution in one dimension, with $T^<[X/t]=t^{k+1} T^<(X)$, the extension of the distribution in the $IR$ domain is given by
\begin{equation} \label{TIR2}
\widetilde T^<_\eta(X)=(-1)^{k}\partial_{X}^{k+1} \left[ \frac{X^{k+1}}{k!} T(X) \mbox{Log} (\tilde \eta X)\right] 
	+ \frac{(-1)^k}{k!} H_k C^k \delta^{(k)}(X)\ ,
\end{equation}
with $\tilde \eta = \eta^2-1$. Note that the surface term in (\ref{TIR2}) does not contribute to the calculation of the amplitude (\ref{cala}) thanks to the presence of the test function. It is however mandatory in order to recover, for instance, the exact propagator of the scalar field in coordinate space, in a perturbative expansion in terms of the mass operator, despite the presence of $IR$ singularities of increasing order \cite{GW_mass,prop1, prop2}.

The usual singular distributions in the IR domain are of the form $T^<(X)=1/X^{k+1}$. In this case,  $\widetilde T^<(X)$ reads, apart from the surface term,
\begin{equation} \label{derlog}
\widetilde T^<(X)=\frac{(-1)^{k}}{k!} \partial_X^{k+1}\mbox{Log} \left(\tilde \eta X \right).
\end{equation}
The derivative of the Logarithmic function should be understood in the sense of distributions, {\it i.e.}
for an arbitrary amplitude involving a continuous function\,$\chi$
\begin{equation}
<\frac{d}{dx}\mbox{Log}(x),\chi>
=\lim_{\epsilon \to 0} \int_\epsilon^\infty \mbox{Log}(x)\left[ -\chi'(x)\right] dx
=\lim_{\epsilon \to 0}\left[ \int_\epsilon^\infty \frac{\chi(x)}{x} dx +\chi(\epsilon)\mbox{Log}(\epsilon) \right] .
\end{equation}
With $\chi(\epsilon)\simeq\chi(0)+\epsilon \chi'(\xi)$, and since $\chi'(\xi)$ is finite, we have
\begin{equation}
<\frac{d}{dx}\mbox{Log}(x),\chi>=\\
\lim_{\epsilon \to 0} \left[ \int_\epsilon^\infty \frac{\chi(x)}{x} dx + \chi(0) \mbox{Log}(\epsilon) \right]\ .
\end{equation}
This is  precisely the definition of the pseudofunction, denoted by Pf, of $1/x$ introduced in Ref.~\cite{schwartz} \footnote{It is also known as the ``{\it finite part}'' (``{\it partie finie''} in french) of a divergent integral, as defined by Hadamard \cite{hada}.}. More generally, we get, for any $k$,
\begin{equation}
\widetilde T^<(X) = \mbox{Pf} \left( \frac{1}{X^{k+1}} \right) . 
\end{equation}
The extension $\widetilde T^<(X)$ differs from the original distribution 
$T^<(X)$ only at the $X=0$ singularity.

\section{Calculation of the singular integrals} \label{RI}
\subsection{Integral in $K^2$} \label{RIK2}
The integral $J_2$ in (\ref{inteb}) can be easily calculated  in $TLRS$. For $m\neq 0$, it is divergent - in the absence of the test functions - only in the $UV$ limit. In the continuum limit, the product $F_\alpha$ of the test functions in (\ref{inteb}) is equivalent, as a partition of unity, to a single test function depending on the dimensionless variable $X={\bf K}^2/m^2$ where ${\bf K}$ is the four-momentum in Euclidian space. We can thus write $J_2$ as
\begin{equation}
J_2=\frac{1}{4\pi^2} \lim_{\alpha \to 1^-} \int_0^{X_{max}} dX \frac{X^2}{(X+a)^3} f_\alpha(X),\nonumber
\end{equation}
with $a=\Delta^2/m^2$. We have chosen here $M_0\equiv m$. 
From the extension of the singular operator (\ref{Tex}) with $k=0$, and after separating out the scale $a$, we  get
\begin{equation}
J_2=\frac{1}{4\pi^2}  \int_0^\infty dX\frac{ \partial}{\partial X}\left[\frac{X X^2}{(X+1)^3}\right] \int_a^{\eta^2} \frac{dt}{t} 
=\frac{1}{4\pi^2} \mbox{Log}\left(\frac{\eta^2}{a}\right).\nonumber
\end{equation}
\subsection{Integral in $K^\mu$}
The calculation of the integral $J_1^\mu$ in (\ref{intec}) should be done with care since the test functions do depend on all the relevant momenta of the system. From the identity
\begin{equation}
\frac{\partial}{\partial {\bf K}_\mu} \frac{1}{({\bf K}^2+\Delta^2)^2}=-4 \frac{{\bf K}^\mu}{({\bf K}^2+\Delta^2)^3},\nonumber
\end{equation}
we can write immediately
\begin{equation}
J_1^\mu=-\frac{1}{4}\lim_{\alpha \to 1^-}\int \frac{d^4 {\bf K}}{(2\pi)^4}\frac{\partial}{\partial {\bf K}_\mu} \left[\frac{1}{({\bf K}^2+\Delta^2)^2} \right]F_\alpha.\nonumber
\end{equation}
By integration by part, the surface term is a 3-dimensional integral orthogonal to the $\mu$-direction. It should be taken at ${\bf K}_\mu \to \pm \infty$. Thanks to the presence of the test functions, this term is identically zero. The remaining integral involves the derivative of $F_\alpha$, with 
\begin{equation}
F_\alpha=f_\alpha\left[\frac{({\bf K}-{\bf Q})^2}{m^2}\right]f_\alpha\left[\frac{({\bf K}-{\bf Q1})^2}{m^2}\right]
f_\alpha\left[\frac{({\bf K}-{\bf Q2})^2}{m^2}\right],\nonumber
\end{equation}
with ${\bf Q1}={\bf Q}-{\bf p_1}$ and ${\bf Q2}={\bf Q}+{\bf p_2}$.  One thus gets
\begin{eqnarray}
J_1^\mu=&&\frac{1}{2m^2}\lim_{\alpha \to 1^-}\int \frac{d^4 {\bf K}}{(2\pi)^4}\frac{1}{({\bf K}^2+\Delta^2)^2}\nonumber\\
\times&& \left[ ({\bf K}-{\bf Q})^\mu f_\alpha^\prime \left[\frac{({\bf K}-{\bf Q})^2}{m^2}\right]f_\alpha\left[\frac{({\bf K}-{\bf Q1})^2}{m^2}\right]f_\alpha\left[\frac{({\bf K}-{\bf Q2})^2}{m^2}\right]\right. \nonumber \\
+&&({\bf K}-{\bf Q1})^\mu f_\alpha \left[\frac{({\bf K}-{\bf Q})^2}{m^2}\right]f_\alpha^\prime\left[\frac{({\bf K}-{\bf Q1})^2}{m^2}\right]f_\alpha\left[\frac{({\bf K}-{\bf Q2})^2}{m^2}\right]\nonumber\\
+&&\left.({\bf K}-{\bf Q2})^\mu f_\alpha \left[\frac{({\bf K}-{\bf Q})^2}{m^2}\right]f_\alpha\left[\frac{({\bf K}-{\bf Q1})^2}{m^2}\right]f_\alpha^\prime\left[\frac{({\bf K}-{\bf Q2})^2}{m^2}\right]\right].\nonumber
\end{eqnarray}
In this equation $f^\prime_\alpha$ denotes $\frac{d}{dX} f_\alpha(X)$. The integral $J_1^\mu$ is a-priori non-zero only in the UV region where $f^\prime\neq 0$. In this region, all test functions are equivalent to $f_\alpha\left[\frac{{\bf K}^2}{m^2}\right]$. By symmetry arguments, the integral over ${ \bf K}_\mu$ is  strictly zero and it remains to calculate
\begin{equation}
J_1^\mu=-\frac{({\bf Q}+{\bf Q1}+{\bf Q2})^\mu}{2m^2} I,\nonumber
\end{equation}
with
\begin{equation}
I=\lim_{\alpha \to 1^-}\int \frac{d^4 {\bf K}}{(2\pi)^4}\\
\frac{1}{({\bf K}^2+\Delta^2)^2}f_\alpha^2\left[\frac{{\bf K}^2}{m^2}\right]f_\alpha^\prime\left[\frac{{\bf K}^2}{m^2}\right]\nonumber
\end{equation}
With the same notations as in \ref{RIK2}, we get
\begin{equation}
I=\frac{1}{3(4\pi)^2}\lim_{\alpha \to 1^-}\int_0^{X_{max}} dX \frac{X}{(X+a)^2}\left[f_\alpha^3(X)\right]^\prime.\nonumber
\end{equation}
By integration by part, the surface term is  zero in the continuum limit $f_\alpha\to 1$, with $X_{max} \to \infty$, and it remains
\begin{equation}
I=-\frac{1}{3(4\pi)^2}\lim_{\alpha \to 1^-}\int_0^{X_{max}} dX \left[\quad\frac{X}{(X+a)^2}\right]^\prime f_\alpha^3(X).\nonumber
\end{equation}
This integral is finite in the absence of the test functions, so that we can also safely take the continuum limit and we finally get
\begin{equation}
J_1^\mu=0. \nonumber
\end{equation}
We recover here rotational invariance. Note that this property is only true in the continuum limit.

\subsection{Integral in $K^\mu K^\nu$}
The calculation of the integral $J_2^{\mu \nu}$ in (\ref{inted}) proceeds similarly. We  start in this case from the identity
\begin{equation}
\frac{\partial}{\partial {\bf K}_\mu} \frac{\partial}{\partial {\bf K}_\nu} \frac{1}{{\bf K}^2+\Delta^2}=
-2\left[\frac{\delta^{\mu \nu}}{({\bf K}^2+\Delta^2)^2} -4\frac{{\bf K}^\mu {\bf K}^\nu}{({\bf K}^2+\Delta^2)^3}\right].\nonumber
\end{equation}
We can thus write
\begin{equation}
J_2^{\mu \nu}=\frac{1}{4} \delta^{\mu\nu}\left[\Delta^2 J_0+J_2\right]+ I^{\mu \nu},\nonumber
\end{equation}
with
\begin{equation}
I^{\mu \nu}=\frac{1}{8} \lim_{\alpha \to 1^-} \int \frac{d^4 {\bf K}}{(2\pi)^4}\frac{\partial}{\partial {\bf K}_\mu} \frac{\partial}{\partial {\bf K}_\nu}\left[\frac{1}{{\bf K}^2+\Delta^2} \right]F_\alpha.\nonumber
\end{equation}
The calculation of $I^{\mu \nu}$ is very similar to the one detailed above for $J_1^{\mu}$. After integration by parts, with the surface terms being zero thanks to the presence of the test functions, we can write
\begin{equation}
I^{\mu \nu}=I^{\mu \nu}_1+I^{\mu \nu}_2+I^{\mu \nu}_3,\nonumber
\end{equation}
where $I^{\mu \nu}_2$ and $I^{\mu \nu}_3$ are deduced from $I^{\mu \nu}_1$ by circular permutation of ${\bf Q}\to {\bf Q_1} \to {\bf Q_2}$, with 
\begin{multline}
I^{\mu \nu}_1=\frac{1}{2m^4}\lim_{\alpha \to 1^-}\int \frac{d^4 {\bf K}}{(2\pi)^4}
\frac{1}{{\bf K}^2+\Delta^2}\left[ \frac{m^2}{2} \delta^{\mu\nu}f_\alpha^2 f_\alpha^\prime+{\bf K}^\mu {\bf K}^\nu\left[ f_\alpha^2 f_\alpha^{\prime \prime}+2f_\alpha(f_\alpha^\prime)^2\right]\right.+\\
\left. {\bf Q}^\mu {\bf Q}^\nu f_\alpha^2f_\alpha^{\prime \prime}+({\bf Q}^\mu {\bf Q_1}^\nu+{\bf Q}^\mu {\bf Q_2}^\nu)  f_\alpha \left(f_\alpha^\prime\right)^2\right].\nonumber
\end{multline}
We removed for simplicity  in this equation the argument $\left[\frac{{\bf K}^2}{m^2}\right]$ for all the test functions and their derivatives. We shall show in the following that the four different integrals entering into the r.h.s. of the above equation, and denoted by $L,M^{\mu\nu},N,O$ respectively, are exactly zero. We first have, apart from uninteresting constant factors and with the same notations as in \ref{RIK2},
\begin{equation}
L=\lim_{\alpha \to 1^-}\int \frac{d^4 {\bf K}}{(2\pi)^4}
\frac{1}{{\bf K}^2+\Delta^2}\left[ f_\alpha^2 f_\alpha^\prime\right] 
=\frac{m^2}{3(4\pi)^2}\lim_{\alpha \to 1^-}\int_0^{X_{max}} dX \frac{X}{X+a}\left( f_\alpha^3\right)^\prime.\nonumber
\end{equation}
It is easy to calculate the integral over X, by integration by part. In the continuum limit, {\it i.e.} with $f_\alpha\to 1$ and $X_{max} \to \infty$, we have
\begin{equation}
L=\frac{m^2}{3(4\pi)^2}\left[ \left.\frac{X}{X+a} \right \vert_0^\infty-\int_0^\infty dX\left(\frac{X}{X+a}\right)^\prime\right]=0.\nonumber
\end{equation}
The integral $M^{\mu\nu}$ writes :
\begin{equation}
M^{\mu\nu}=\lim_{\alpha \to 1^-}\int \frac{d^4 {\bf K}}{(2\pi)^4}
\frac{{\bf K}^\mu {\bf K}^\nu}{{\bf K}^2+\Delta^2}\left[ f_\alpha^2 f_\alpha^\prime\right]^\prime .\nonumber
\end{equation}
Since in the asymptotic region where $f_\alpha^\prime$ is non zero the test functions depend only on ${\bf K}^2$, we have immediately, from rotational invariance at $D=4$
\begin{eqnarray}
M^{\mu\nu}&=&\frac{1}{4}\delta^{\mu\nu}\lim_{\alpha \to 1^-}\int \frac{d^4 {\bf K}}{(2\pi)^4}
\frac{{\bf K}^2}{{\bf K}^2+\Delta^2}\left[ f_\alpha^2 f_\alpha^\prime\right]^\prime \nonumber \\
&=&\frac{m^4}{4(4\pi)^2}\delta^{\mu\nu}\lim_{\alpha \to 1^-}\int_0^{X_{max}} dX\frac{X^2}{X+a}\left[ f_\alpha^2 f_\alpha^\prime\right]^\prime.\nonumber
\end{eqnarray}
By integration by part, we get
\begin{equation}
M^{\mu\nu}=\frac{m^4}{4(4\pi)^2}\delta^{\mu\nu}\lim_{\alpha \to 1^-}\left[ \left.\frac{X^2}{X+a} f_\alpha^2 f_\alpha^\prime\right \vert_0^{X_{max}}
-\int_0^{X_{max}}dX\left(\frac{X^2}{X+a}\right)^\prime f_\alpha^2 f_\alpha^\prime\right].\nonumber
\end{equation}
In the above expression, the surface term is identically zero in the continuum limit, since in this limit the derivative of the test function tends to zero  more rapidly than any power of X (super-regular function). The last integral can be calculated in the same way as $L$, with the result
\begin{equation}
M^{\mu\nu}=-\frac{m^4}{12(4\pi)^2}\delta^{\mu\nu}\left[ \left.\frac{X^2+2a X}{(X+a)^2} \right \vert_0^\infty \right. 
\left.-\int_0^\infty dX\left(\frac{X^2+2a X}{(X+a)^2}\right)^\prime\right] =0.\nonumber
\end{equation}
The integral $N$ can be calculated after the replacement $f_\alpha^2 f_\alpha^{\prime\prime}=\left(f_\alpha^2 f_\alpha^\prime\right)^\prime-2f_\alpha (f_\alpha^\prime)^2$. The second term of the above equation will thus be incorporated in the calculation of $O$. Keeping only the first term, we get
\begin{equation}
N=\lim_{\alpha \to 1^-}\int \frac{d^4 {\bf K}}{(2\pi)^4}
\frac{1}{{\bf K}^2+\Delta^2}\left[ f_\alpha^2 f_\alpha^\prime\right]^\prime 
=\frac{m^2}{(4\pi)^2}\lim_{\alpha \to 1^-}\int_0^{X_{max}} dX\frac{X}{X+a}\left[ f_\alpha^2 f_\alpha^\prime\right]^\prime.\nonumber
\end{equation}
After integration by part, the surface term is identically zero in the continuum limit, like in the calculation of $M^{\mu\nu}$. It remains
\begin{eqnarray}
N&=&-\frac{m^2}{3(4\pi)^2}\lim_{\alpha \to 1^-}\int_0^{X_{max}} dX
\left(\frac{X}{X+a}\right)^\prime \left(f_\alpha^3\right)^\prime \nonumber \\
&=&-\frac{a m^2}{3(4\pi)^2}\lim_{\alpha \to 1^-}\left[\left.\frac{1}{(X+a)^2} f_\alpha^3\right\vert_0^{X_{max}}
-\int_0^{X_{max}} dX\left[\frac{1}{(X+a)^2}\right]^\prime f_\alpha^3\right].\nonumber
\end{eqnarray}
In the continuum limit, we get immediately
\begin{equation}
N=0\nonumber
\end{equation}
The last integral we have to calculate is given by
\begin{eqnarray}
P&=&\lim_{\alpha \to 1^-}\int \frac{d^4 {\bf K}}{(2\pi)^4}
\frac{1}{{\bf K}^2+\Delta^2}\left( f_\alpha^\prime\right)^2 f_\alpha 
=\frac{m^2}{2(4\pi)^2}\lim_{\alpha \to 1^-}\int_0^{X_{max}} dX \frac{X}{X+a} \left( f_\alpha^2\right)^\prime f_\alpha\nonumber \\
&=&\frac{m^2}{2(4\pi)^2}\lim_{\alpha \to 1^-}\left[\left.\frac{X}{X+a} f_\alpha^2 f_\alpha^\prime\right\vert_0^{X_{max}}
-\int_0^{X_{max}} dX\left[\frac{X}{X+a} f_\alpha^\prime\right]^\prime f_\alpha^2\right].\nonumber
\end{eqnarray}
From the properties of the test functions, and its derivative, in the continuum limit, we thus get also
\begin{equation}
P=0\nonumber
\end{equation}
To summarise, we finally have
\begin{equation}
I^{\mu \nu}=0\ \ \ \mbox{and}\ \ \ 
J_2^{\mu \nu}=\frac{1}{4} \delta^{\mu\nu}\left[\Delta^2 J_0+J_2\right].\nonumber
\end{equation}


\begin{thebibliography}{99}
\bibitem{ryder}
L.H. Ryder, ``{\it Quantum field theory}'', Ed. Cambridge University Press, (1985).
\bibitem{pauli}
C. Itzykson and J.B. Zuber, ``{\it Quantum Field Theory}'', Mc Graw Hill Eds, (1980).
\bibitem{collins}
J. Collins, `` {\it Renormalization}'',  Ed. Cambridge University Press, (1984).
\bibitem{EG}
H. Epstein and V. Glaser, Ann. Inst. Henri Poincar\'e, {\bf XIX A} (1973) 211.
\bibitem{Scharf} 
G. Scharf, ``{\it Finite QED: the causal approach}'', Springer Verlag (1995).
\bibitem{prop1}
J.M. Gracia-Bondia, Math. Phys. Anal. Geom. {\bf 6} (2003) 59;
\bibitem{prop2}
J.M. Gracia-Bondia and S. Lazzarini, J. Math. Phys. {\bf 44} (2003) 3863.
\bibitem{GW_mass}
P. Grang\'e and E. Werner, Nucl. Phys. (Proc. Suppl.) {\bf B161}, 75 (2006).
\bibitem{GW}
P. Grang\'e and E. Werner, J.of Phys. A: Math. Theor. {\bf 44}, 385402 (2011).
\bibitem{bogo}
N.N. Bogoliubov, Doklady USSR Acad. Sci. {\bf 82}, 217 (1952).
\bibitem{bosh}
N.N. Bogoliubov and D.V. Shirkov,
{\it  An introduction to the Theory of Quantized Fields}, New York, J. Wiley \& Sons, Publishers, Inc.,  (1959). 
\bibitem{wigh}
A.S. Wightman, Phys. Rev. {\bf 101}, 860 (1956).
\bibitem{haag}
R. Haag ``{\it Local Quantum Physics: Fields, Particules, Algebras}'', (Berlin, Heidelberg, New York: {\it Texts and Monographs in Physics, 2nd edition} Springer-Verlag) (1996).
\bibitem{schw}
S. Schweber, {\it An introduction to relativistic quantum field theory, Sec.18}, Harper and Row, N.Y., Eds, (1964).
\bibitem{tkac}
F.V. Tkachov, ``{\it talk at the Bogolyubov Conference on Problems of Theoretical and Mathematical Physics}'', Moscow-Dubna-Kyiv, september 1999 (ArXiv hep-th/9911236).
\bibitem{stue}
E.C.G. Stueckelberg and A. Petermann, Helv. Phys. Acta {\bf 26}, 498 (1953).
\bibitem{HORM1}
L. Hormander, ``{\it Linear Partial Differential Equation}'', (Berlin: Springer-Verlag)(1969).
\bibitem{HORM2}
L. Hormander, ``{\it The Analysis of Linear Partial Differential Operators}'', (Berlin: Springer-Verlag)(1986).
\bibitem{LFD}
P. Grang\'e, J.-F. Mathiot, B. Mutet, E. Werner, Phys. Rev. {\bf D80}, 105012 (2009).
\bibitem{LFD_PV}
P. Grang\'e, J.-F. Mathiot, B. Mutet, E. Werner, Phys. Rev. {\bf D82}, 025012 (2010).
\bibitem{GW_gauge}
B. Mutet, P. Grang\'e and E. Werner, J.of Phys. A: Math. Theor. {\bf 45}, 315401 (2012).
\bibitem{higgs}
P. Grang\'e, J.-F. Mathiot, B. Mutet, E. Werner,  Phys. Rev. {\bf D88}, 125015 (2013).
\bibitem{schwartz}
L. Schwartz, ``{\it Th\'eorie des distributions}'', Hermann, Paris 1966 ($1^{st}$ ed.:1950-1951);
 ``{\it Mathematics for the physical sciences}'', Dover Publication (1966).
\bibitem{gene}
S.L. Sobolev, Mat. Sbornik 1(43) 39 (1935). 
\bibitem{Integ}
D.L. Cohn  {\it Measure Theory} Birkh$\ddot{a}$user, Boston (1980). 
\bibitem{jfm}
J.-F. Mathiot, Int. J. Mod. Phys. {\bf A33} 1830024 (2018)
\bibitem{PU1}
J. Dieudonn\'e, Math. Pure Appl. {\bf} 23, 65 (1944).
\bibitem{PU2}
A.H. Stone, Bull. Am. Math. Soc. {\bf 54}, 977 (1948).
\bibitem{PU3}
Nakahara M 1990 Geometry, Topology and Physics (Graduate Student Series in Physics) (Bristol: Institute of
Physics Publishing)
\bibitem{hada}
J. Hadamard, ``{\it Le probl\`eme de Cauchy et les \'equations aux d\'eriv\'ees partielles lin\'eaires hyperboliques}'', Paris, Hermann (1932).
\bibitem{Hgg}
J. Ellis, M.K. Gaillard, and D.V. Nanopoulos, Nucl. Phys. {\bf B106}, 292 (1976).
\bibitem{UlfM}
J. Gegelia and Ulf-G. Meissner, Nucl. Phys. {\bf B934}, 1 (2018).
\bibitem{WWW}
R. Gastmans, S.L. Wu and T.T. Wu, Int. J. Mod. Phys. {\bf A30}, 1550200 (2015).
\end{thebibliography}
\end{document}